\documentclass[aps,prx,twocolumn,showkeys,floatfix,longbibliography]{revtex4-2}
\usepackage[T1]{fontenc}
\usepackage{tabularx}
\usepackage{dcolumn}
\usepackage{amsmath,amssymb,amsfonts,bm,dsfont,units}
\usepackage{hyperref}
\hypersetup{colorlinks=true,citecolor=blue,linkcolor=blue,urlcolor=blue,pdfstartview=FitH}
\usepackage{appendix} 
\usepackage{braket}
\usepackage{latexsym}
\usepackage{multirow}
\usepackage{diagbox}
\usepackage[normalem]{ulem} 
\usepackage[caption=false]{subfig} 
\usepackage{xcolor} 
\usepackage{array}
\usepackage{graphicx} 
\usepackage{hhline,booktabs}
\usepackage{url}
\usepackage{makecell}
\usepackage{wasysym}  
\usepackage{tikz}
\usetikzlibrary{shapes,arrows,positioning} 

 \usepackage{times} 
 \definecolor{darkblue}{HTML}{004D6B}
 \definecolor{darkred}{HTML}{8c1515}
 \definecolor{darkgreen}{HTML}{006400}
 \hypersetup{
     pdftitle={MUD},
 	colorlinks=true,
 	urlcolor=darkblue,
 	citecolor=darkred,
 	linkcolor=darkred,
 	breaklinks
 }
 \pagecolor{white}

\newcommand{\vect}[1]{\boldsymbol{#1}}
\renewcommand{\d}{L}

\begin{document}

\title{Optimal Decoding for Measurement-Based GHZ State Preparation: \\  The Maximum-Utility Decoder}

\author{Misha Yutushui}
\affiliation{Institute for Theoretical Physics, University of Cologne, Zülpicher Straße 77, 50937 Cologne, Germany}

\author{Theo Haas}
\affiliation{Institute for Theoretical Physics, University of Cologne, Zülpicher Straße 77, 50937 Cologne, Germany}

\author{Simon Trebst}
\affiliation{Institute for Theoretical Physics, University of Cologne, Zülpicher Straße 77, 50937 Cologne, Germany}

\begin{abstract}  
The meticulous preparation of macroscopic Greenberger-Horne-Zeilinger (GHZ) states provides a foundational resource for quantum technologies such as metrology, cryptography, and fault-tolerant codes.
While state-of-the-art measurement-based protocols offer efficient low-depth execution, their performance can be bottlenecked by conventional decoders, such as minimum weight perfect matching (MWPM) or even maximum-likelihood decoding (MLD), which optimize for {\sl binary} logical recovery and fail to maximize the {\sl continuous} long-range order characteristic of a GHZ state for two-dimensional geometries.
Here we overcome this limitation by framing the decoding problem as minimum Bayesian risk inference, introducing a general paradigm that maximizes the expected {\sl utility} of the decoded state.
Implementing this maximum-utility approach, we construct an algorithm that achieves the highest possible per-shot decoded quantum order
and thereby establish an optimal decoding strategy for measurement-based GHZ state preparation. To improve its computational efficiency, we design a scalable two-stage decoder, which first encodes the syndromes into the edge weights of MWPM and then refines the result with a convolutional neural network trained to maximize the expected utility, at a fraction of the cost of the optimal decoder. Remarkably, we find that the first stage alone---which makes the matching aware of the gauge choice at no cost beyond bare MWPM---already performs near-optimally up to the largest sizes we study, $N=256\times256$, closing up to $87\%$ of the gap between the bare-MWPM and optimal decoding thresholds.
Generalizing MWPM and MLD, the maximum-utility decoder (MUD) establishes a versatile framework that can be explicitly tailored to the operational demands of specific experiments by redefining the utility function.
\end{abstract}

\maketitle

\section{Introduction}
The Greenberger-Horne-Zeilinger (GHZ) state is a paradigmatic example of a quantum state with many-qubit entanglement~\cite{Greenberger_GHZ_1989} that acts as an indispensable resource across quantum information science. Its macroscopic superposition and perfect multi-qubit correlations enable metrological sensing at the Heisenberg limit~\cite{Bollinger_Optimal_1996,Leibfried_Toward_2004,Giovannetti_Quantum_2004,Giovannetti_Advances_2011,Marciniak_Optimal_2022} and ensure robust security in cryptographic secret sharing~\cite{Hillery_secret_1999,Cleve_How_1999,Chen_Experimental_2005}. 
GHZ states also provide the essential entanglement infrastructure for distributed quantum networking~\cite{Grover_quantum_1997,Cirac_Distributed_1999,Bone_Protocols_2020}, and serve as vital ``cat states" in repetition-code  architectures~\cite{Shor_Scheme_1995,Shor_Fault_1996,Guillaud_Repetition_2019,Takeda_Quantum_2022,Bergamaschi_Fault_tolerant_2025,Putterman_Hardware_2025}.

Deterministically preparing macroscopic long-range entangled states, such as the GHZ state, remains a challenging problem on near-term hardware. In local unitary circuits, the entanglement growth is constrained by Lieb-Robinson bounds~\cite{Bravyi_Lieb_2006, Hastings_Locality_2010}, i.e., the circuit depth scales linearly with the number of qubits. On current noisy intermediate-scale quantum (NISQ) devices~\cite{Preskill_Quantum_2018}, this extended circuit execution time still remains a critical liability; the long execution times subject the physical qubits to continuous environmental noise, potentially leading to catastrophic decoherence before the entanglement can be safely utilized.

\begin{figure}[b]
    \centering
    \includegraphics{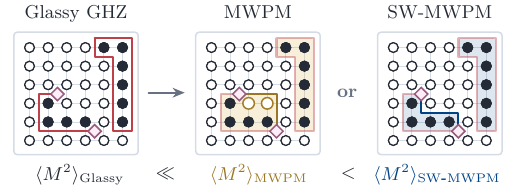}
    \caption{\textbf{Glassy GHZ state decoding.} 
    Shown are schematic data-qubit configurations (white and black circles) along with measured syndromes (domain walls) indicated by red lines.
    If the lines form a closed loop, they enclose a uniform domain, while in the case of an open loop, they terminate on fluxes marked by  
    pink diamonds, leaving the identification of domains to a decoder.
    The two panels on the right show two possible decoding outcomes that aim to close the open loop configurations employing a minimum weight 
    perfect matching (MWPM) or a syndrome-weighted SW-MWPM approach, with the shaded regions indicating which qubits will be flipped in
    the correction step. The quality of a decoder can be quantified via the squared magnetization $\langle \hat{M}^2\rangle$,
    a continuous variable that quantifies the amount of long-range quantum correlations.
	}
	\label{fig:decoding_problem}	
\end{figure}

Measurement-based protocols~\cite{Raussendorf_One_way_2001} overcome this limitation by employing mid-circuit measurements to bypass Lieb-Robinson bounds and prepare long-range entangled states in constant depth~\cite{Verresen_Efficiently_2021,Tsung_Measurement_2022}. 
However, the immediate post-measurement system is not the target state; the joint parity-check measurements naturally collapse the system into a randomized, ``glassy'' GHZ state characterized by domains of opposite orientation~\cite{Zhu_Nishimori_2023,Lee_decoding_2022}. Consequently, the success of the protocol relies on a classical decoder to process the measurement outcomes, infer the domain boundaries, and apply restorative feedforward corrections. While a perfect decoder operating on noiseless hardware would restore a pristine GHZ state, near-term hardware suffers from imprecise gates and readout noise~\cite{Chen_Nishimori_2025}. Under these realistic conditions, it is impossible to deterministically recover the maximally correlated state. Nevertheless, the noisy protocol prepares a finite amount of long-range entanglement, and the target of an optimal decoder shifts to restoring the macroscopic order.

\begin{figure*}[t]
    \centering
    \includegraphics{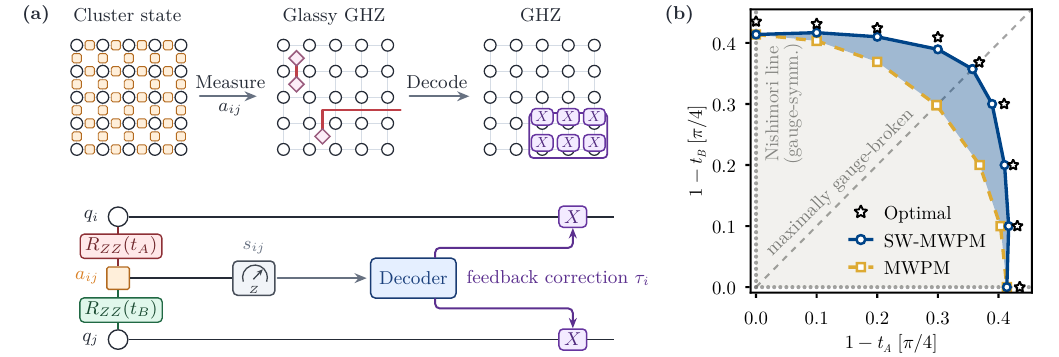}
     \caption{{\bf Measurement-based GHZ state preparation and decoding.} (a) The measurement-based preparation and decoding protocol starts from a two-dimensional Lieb-lattice structure, with data qubits $q_i$ (white dots) and auxiliary stabilizer qubits $a_{ij}$ (orange squares). Measuring the stabilizers produces a glassy GHZ state whose outcomes define syndrome variables $s_{ij}$ on the links. Links with outcome $s_{ij}=-1$ are drawn as red lines; connected segments form syndrome strings, whose open endpoints define flux excitations marked by pink diamonds. The syndromes are then passed to a decoder, which returns a feedback correction $\tau_i$; the resulting correction pattern is shown in purple. (b) Decoding thresholds in the $(t_A,t_B)$ plane are shown for MWPM (yellow squares), syndrome-weighted MWPM (blue circles), and the theoretical optimal limit (star markers). The dotted line marks the Nishimori line ($t_A=\pi/4$ or $t_B=\pi/4$), where the decoding problem exhibits a local $\mathbb{Z}_2$  gauge symmetry, while the dashed diagonal line ($t_A=t_B$) marks the case of maximal gauge-symmetry breaking. The threshold estimates are obtained by a finite-size scaling analysis of the Binder cumulant of the squared magnetization $\langle \hat{M}^2\rangle$, as shown in Fig.~\ref{fig:decoded_M}.}
      \label{fig:setup}
\end{figure*}

The objective of decoding a GHZ state fundamentally differs from decoding standard topological codes~\cite{Dennis_Topological_2002,Fowler_Surface_2012,deMarti_Decoding_2024}. For the toric code, the precise microscopic path of an applied correction is unimportant, provided it achieves the correct logical operation. By contrast, when preparing a GHZ state, local deviations directly impact the physical long-range order. Any discrepancy between the true error and the applied correction leaves behind an uncorrected domain that reduces macroscopic correlations; see Fig.~\ref{fig:decoding_problem}.

Minimum weight perfect matching (MWPM)~\cite{kolmogorov2009blossom, higgott2025sparse} is a natural heuristic for this task, producing the shortest, most likely error configuration. However, MWPM is fundamentally suboptimal because it cannot differentiate between highly degenerate shortest paths that yield different amounts of long-range order, see Fig.~\ref{fig:decoding_problem}. In more technical terms, MWPM infers the correction solely from {\sl gauge-invariant} data---the endpoints of the syndrome chains. This is justified only when the decoding problem maps to a Nishimori line, where the system indeed exhibits a local $\mathbb{Z}_2$ gauge symmetry \cite{Nishimori_1980}. Away from the Nishimori line, the gauge symmetry is broken, and the precise configuration of syndromes becomes actionable information ignored by the MWPM decoder, making it suboptimal.

To overcome this limitation, we introduce a decoding paradigm where success is measured by a continuous {\sl utility function} and
which guides the implementation of an optimal decoder for a specific task. 
 Rooted in minimum Bayesian risk inference \cite{berger1985decision}, the decoder maximizing the expected utility---maximum-utility decoder (MUD)---presents a generalized architecture that naturally reduces to MWPM or maximum likelihood decoding (MLD) when the utility function is tailored for standard surface-code decoding~\cite{Dennis_Topological_2002,Bravyi_Efficient_2014}. For GHZ state preparation, we define the utility as the squared magnetization---a direct measure of long-range order. Specifically, the decoder searches the landscape of possible corrections and selects the configuration that maximizes the expected macroscopic correlations. By natively processing the full geometric information of the syndromes rather than just their endpoints, the MUD systematically extracts the highest-fidelity GHZ state from the noisy hardware, establishing an optimal decoding paradigm for measurement-based state preparation.

\section{Surface-code decoding}
To establish MUD as a generalized decoding paradigm, we first frame standard topological code decoding within a Bayesian decision-making framework. Consider the toric code subjected to bit-flip noise. The error configuration $\vect{e}$ occurs on the edge qubits with probability $P(\vect{e})$. Measuring the plaquette syndromes $\vect{f}=\partial \vect{e}$ identifies parity violations, where $\partial$ denotes the dual boundary operator that maps the dual edge chain $\vect{e}$ to its endpoints. Based on the observed syndromes $\vect{f}$, a decoder must suggest a correction chain $\vect{c}$ such that $\partial \vect{c} = \vect{f}$.

Within decision theory, a decoder's objective can be universally defined by a utility function $\mathcal{U}(\vect{c},\vect{e})$ that quantifies the success of a correction $\vect{c}$ given a true error $\vect{e}$. The optimal inference problem is equivalent to maximizing the expected utility given the syndrome information
\begin{align}
    \text{MUD: }\vect{f}\to \vect{c} = \mathop{\operatorname{argmax}}\limits_{\vect{c}} \sum_{\vect{e}}\mathcal{U}(\vect{c},\vect{e})P(\vect{e}|\vect{f}) \,,
    \label{eq:MUD}
\end{align}
where $P(\vect{e}|\vect{f}) = P(\vect{e})\delta(\vect{f}-\partial \vect{e})/ P(\vect{f})$ is the {\sl conditional} error probability. Standard decoding algorithms merely correspond to different choices of the utility function
\begin{align}\label{eq.U_surface}
\mathcal{U}(\vect{c},\vect{e}) = 
\begin{cases}
	\delta(\vect{e}-\vect{c})\,,		&	\text{for MWPM}\,,\\
	\delta([\vect{e}]-[\vect{c}])\,,	&	\text{for MLD}\,,\\
\end{cases}
\end{align}
where $[\cdot]$ denotes the homology class of the chain. By penalizing any microscopic mismatch, MWPM finds the most likely error configuration given the syndrome. In contrast, MLD prioritizes logical fidelity instead of microscopic accuracy by recognizing that any error in the same homology class as the correction forms a trivial loop that does not change the logical information. Because all errors within this class yield identical logical outcomes, the MLD utility function treats the entire class {\sl uniformly}, maximizing the aggregate probability of a successful recovery.

In the topological cases, the utility landscape is {\sl binary}, $\mathcal{U}(\vect{c},\vect{e})=0$ or $1$, and the utility function merely defines certain sets of errors. Conversely, for non-topological states, a utility function is needed as the quality of the decoded state is characterized by a {\sl continuous} order parameter, e.g., the decoded magnetization squared $\mathcal{U}_\text{GHZ}(\vect{c},\vect{e})=\langle \hat{M}^2\rangle$ for the GHZ state, which is closely related to the fidelity~\footnote{A shot fidelity is ${\cal F}=1$ for a perfectly decoded shot ($\langle \hat{M}^2\rangle=1$) and ${\cal F}=0$ otherwise. It corresponds to the utility function $\mathcal{U}_{\cal F}(\vect{\tau},\vect{\sigma}) = \theta\!\left(\frac{1}{N}\sum_{i}\sigma_i \tau_i-1\right)$, whose Heaviside nonlinearity renders it analytically intractable.}.
We now demonstrate how MUD improves the performance of the measurement-based GHZ-state preparation protocol. 

\section{Measurement-based protocol}

To prepare a macroscopic GHZ state, we employ a constant-depth measurement protocol on a square Lieb lattice geometry; see Fig.~\ref{fig:setup}(a). The procedure first entangles data qubits on the vertices with auxiliary stabilizer qubits on the lattice edges. Measuring these auxiliary qubits implements a joint parity check of the two adjacent data qubits that yields a classical outcome $s_{ij} = \pm 1$, which we treat as a syndrome configuration. The latter can be viewed as a chain of dual edges; see the middle panel of Fig.~\ref{fig:setup}(a).  Even in a noiseless scenario, this measurement projects the data qubits into a ``glassy'' GHZ state, where domains of opposite orientation are separated by closed dual loops of $s_{ij}=-1$, corresponding to ``antiferromagnetic'' domain walls. 
Such closed-loop configurations can be decoded deterministically by flipping one type of domains to produce a pristine GHZ state;
this is the classical decoding step that every measurement-based state preparation protocol requires.

However, realistic hardware suffers from imprecise operations, which we model here by replacing ideal entangling gates with asymmetric coherent rotations $R_{ZZ}(2t_{A/B}) = e^{-it_{A/B} Z Z}$ applied between sub-lattices $A$ and $B$; see Fig.~\ref{fig:setup}. With these imperfections,  {\sl open} antiferromagnetic chains are produced $\partial \vect{s}=\vect{f}$ with endpoints at gauge-invariant plaquette fluxes $f_\square = -1$ that cannot be eliminated by bit-flip corrections,
see Fig.~\ref{fig:decoding_problem}. Observing the syndrome chain $\vect{s}$ implies that the data qubits are weakly measured and their wavefunction is partially projected with the Kraus operator
\begin{align}\label{eq.psi_s}
    |\Psi(\vect{s})\rangle \propto
    e^{\frac{\beta}{2} \sum_{\langle ij\rangle}J_{s_{ij}} Z_i Z_j}|+\rangle^{\otimes N}\,,
\end{align} 
where the summation is over neighboring vertices and the effective bond couplings satisfy 
$\tanh \frac{\beta J_{s}}{2} = s [\tan t_B] [\tan t_A]^{-s}$ for $t_A\geq t_B$~\cite{Zhu_Nishimori_2023}. 
(The case of $t_A <  t_B$ follows from  symmetry by swapping $t_A\leftrightarrow t_B$.) 
The probability of measuring a syndrome configuration $\vect{s}$ is given by the partition function of the 2D random-bond Ising model (RBIM) $P(\vect{s})=Z_\text{RBIM}[\vect{s}] \propto \sum_{\vect{\sigma}} e^{-\beta H(\vect{s})}$  with Hamiltonian
\begin{align}\label{eq.H_s}
    H(\vect{s}) = -\sum_{\langle ij\rangle}\left( J_{s_{ij}}\sigma_{i}\sigma_j +h s_{ij}\right) \,,
\end{align}
where $\beta h = \frac{1}{2}\log|\tan(t_A+t_B)\tan(t_A-t_B)|$ and $\sigma_i=\pm1$ can be viewed as the $Z_i$ eigenvalue of the vertex qubits. 

\subsection{Domain walls and open loops}

Prior to syndrome measurements, the protocol prepares a quantum state with a joint spin-syndrome probability distribution
\begin{align}\label{eq.P_s_sigma}
    P(\vect{\sigma},\vect{s}) =\frac{1}{2^N} \prod_{\langle ij \rangle} \frac{1-s_{ij} \cos 2 (\sigma_i t_A + \sigma_j t_B)}{2} \,.
\end{align}
If we marginalize over the spin configurations $\vect{\sigma}$, the distribution reduces to the  RBIM partition function $\sum_{\vect{\sigma}}P(\vect{\sigma},\vect{s})=Z_\text{RBIM}[\vect{s}]$ with the Hamiltonian in Eq.~\eqref{eq.H_s}. Conversely, marginalizing over $\vect{s}$ yields a uniform distribution of vertex spins, $P(\vect{\sigma})=\frac{1}{2^N}$. Hence, prior to the correction stage, the data qubits exhibit no long-range order. 

The conditional probability $P(\vect{s}|\vect{\sigma})$ implies that the relative alignment of spins $\sigma_i\sigma_j=\pm1$ biases the syndrome measurement toward $s_{ij} = \sigma_i\sigma_j$. Consequently, a typical configuration $\vect{s}$ contains a large number of loops enclosing domains of opposite orientation. 

Concurrently, the gate imperfections $t_A,t_B<\pi/4$ introduce {\sl open syndrome chains}, as illustrated in Fig.~\ref{fig:decoding_problem}. By analogy with the Ising model, we say that a plaquette contains a flux if it is the endpoint of the dual syndrome chain, i.e., plaquettes with an odd number of antiferromagnetic bonds $f_{\square} = \prod_{\langle ij\rangle\in \square} s_{ij} = -1$. These fluxes are gauge-invariant objects in the sense that they cannot be eliminated by bit flips. In contrast to domain walls that reduce long-range order but do not affect long-range entanglement, fluxes are detrimental to both. The structure of the fluxes simplifies along the symmetric Nishimori line, where the syndrome paths become gauge-redundant.

\begin{figure*}[t]
    \centering
    \includegraphics{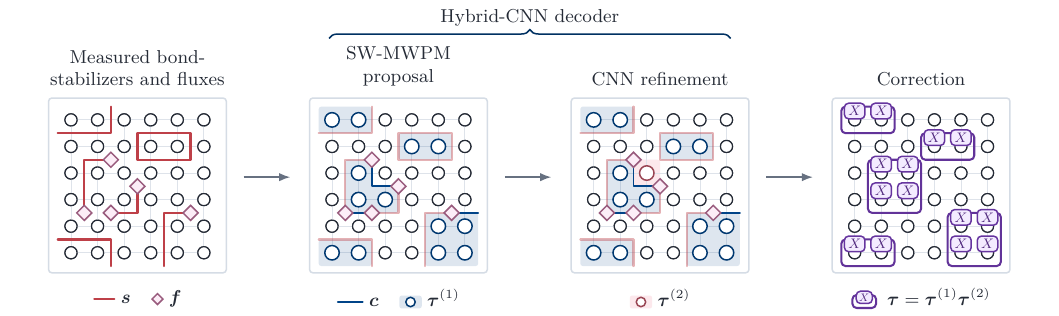}
    \caption{{\bf Schematic decoding pipeline for the hybrid-CNN decoder.} Measuring the edge stabilizers yields the syndrome $\vect{s}$ depicted as red {\sl dual} edges, whose boundary (endpoints) $\vect{f}=\partial\vect{s}$ locates the gauge-invariant plaquette fluxes (magenta). The syndrome-weighted MWPM first proposes a baseline correction $\vect{c}$ (blue), arbitrarily chosen from a number of chains with minimum weight but not necessarily one with the largest recovered long-range order. The CNN then refines the given  $\vect{\tau}^{(1)}$ (enclosed in $\vect{s}\vect{c}=\partial \vect{\tau}^{(1)}$) by proposing a calibration $\vect{\tau}^{(2)}$.  Finally, the correction applies a Pauli-$X$ on a combination of data qubits $\vect{\tau}=\vect{\tau}^{(1)}\vect{\tau}^{(2)}$.
	}    
	\label{fig:decoder}
\end{figure*}

\subsection{Nishimori line}
On the Nishimori line ($t_A = \pi/4$ or $t_B = \pi/4$), the couplings become symmetric $\beta J_{+} = -\beta J_{-}$ and $\beta h=0$. A bit-flip $X_{i_0}$ reverses the signs of all edge syndromes $s_{i_0j}$ adjacent to vertex $i_0$
\begin{align}\label{eq.commutator}
X_{i_0} e^{\frac{\beta J}{2} s_{i_0j} Z_{i_0} Z_j} = e^{\frac{\beta J}{2}(-s_{i_0j}) Z_{i_0} Z_j} X_{i_0} \,,
\end{align}
where we used  $J_{s_{ij}} = s_{ij} J$ on the Nishimori line. The bit-flip correction $X(\vect{\tau}) = \prod_{i\in \vect{\tau}}X_i$ parametrized by $\tau_i=\pm1$, with $i\in \vect{\tau}$ if $\tau_i=-1$, deforms the syndrome chain
\begin{align}\label{eq.transformed_psi}
    X(\vect{\tau})\;
    |\Psi(\vect{s})\rangle~=~|\Psi(\vect{s}\partial \vect{\tau})\rangle=~|\Psi(\vect{c})\rangle \,,
\end{align}
where $ c_{ij} =  s_{ij}\tau_i\tau_j$ is the edge chain deformed from the syndrome chain $\vect{s}$ by the boundary $\partial\vect{\tau}$. This gauge transformation effectively erases the information about the microscopic path of $\vect{s}$ from the corrected wavefunction. Consequently, the position of the fluxes $\vect{f} = \partial \vect{s}$ represents the only non-trivial information available to a decoder.

\subsection{Away from the Nishimori line}

For arbitrary gate parameters $t_A, t_B < \pi/4$, the local $\mathbb{Z}_2$ gauge symmetry is explicitly broken. Ferromagnetic and antiferromagnetic bonds have asymmetric contributions ($\beta J_+ \neq -\beta J_-$), and Eqs.~\eqref{eq.commutator} and \eqref{eq.transformed_psi} no longer hold. The bit-flip correction 
\begin{align}\label{eq.transformed_psi_away_nishi}
X(\vect{\tau})|\Psi(\vect{s})\rangle~\propto~|\Psi(\vect{s},\vect{\tau})\rangle=
    \prod_{\langle ij\rangle\in \partial \vect{\tau}}e^{-\beta  \Delta J Z_iZ_j} |\Psi(\vect{s}\partial \vect{\tau})\rangle
\end{align}
does not yield a wavefunction of the form in Eq.~\eqref{eq.psi_s} due to a non-trivial contribution for $\Delta J\equiv \frac{J_++J_-}{2}\neq0$. The operator acting on the boundary $\partial \vect{\tau}$ reflects this explicit symmetry breaking. Consequently, the corrected quantum state depends on both the original syndrome chain $\vect{s}$ and the deformed chain $\vect{c} = \vect{s} \partial \vect{\tau}$. Identifying the optimal correction, therefore, requires a decoding procedure that accounts for the complete microscopic syndrome path rather than just its boundary $\vect{f} = \partial \vect{s}$.

\section{Decoding}

To restore macroscopic order, the decoder processes the observed syndromes $\vect{s}$ and outputs a correction on a set of vertices $\vect{\tau}$. Applying the bit-flip correction $X(\vect{\tau})$ results in the decoded state Eq.~\eqref{eq.transformed_psi_away_nishi}, whose squared magnetization is
\begin{align}\label{eq:M2_sc}
    \langle \hat{M}^2 \rangle_{\vect{s},\vect{\tau}} = \frac{\sum_{\vect{\sigma}} M^2 e^{-\beta H(\vect{s},\vect{\tau})}}{ \tilde{Z}[\vect{s},\vect{\tau}]} \,,
\end{align}
with the magnetization $M=\sum_i\sigma_i/N$ and the transformed Hamiltonian 
\begin{align}\label{eq:H_sc}
    H(\vect{s},\vect{\tau}) =  - \sum_{\langle ij\rangle} J_{s_{ij}} \sigma_i\sigma_j\tau_i\tau_j \,.
\end{align}
This quantum expectation value can be written as a classical a posteriori expectation value, 
\begin{align}
    \langle \hat{M}^2 \rangle_{\vect{s},\vect{\tau}} = \sum_{\vect{\sigma}} \mathcal{U}(\vect{\tau},\vect{\sigma}) P(\vect{\sigma}|\vect{s}) \,,
    \\
    \mathcal{U}(\vect{\tau},\vect{\sigma}) = \left(\frac{1}{N}\sum_{i}\sigma_i \tau_i\right)^2 \,,
    \label{eq:utility_function}
\end{align}
derived explicitly in Appendix~\ref{app.MUD}. Thus, the value $\langle \hat{M}^2 \rangle_{\vect{s},\vect{\tau}}$ is precisely the expected utility of the correction $\vect{\tau}$ given the underlying spins $\vect{\sigma}$ that the optimal decoder seeks to maximize. Notice that compared to the surface code, the syndromes $\vect{s}$ here live on edges and the `noise' induces vertex flips $\vect{\sigma}$; thus, the utility function is a function of vertex variables $\vect{\sigma}$ and $\vect{\tau}$, cf. Eq.~\eqref{eq.U_surface} for the surface code. 

\subsection{Optimal decoder}

The expected utility $\langle \hat{M}^2\rangle_{\vect{s},\vect{\tau}}$ can be expressed in terms of the spin-spin correlation function $C_{ij}(\vect{s})=\langle \sigma_i\sigma_j\rangle_{\vect{s}}$ as
\begin{align}
    \langle \hat{M}^2 \rangle_{\vect{s},\vect{\tau}} = \frac{1}{N^2}\sum_{i,j} C_{ij}(\vect{s}) \tau_i\tau_j \,.
\end{align}
Therefore, maximizing the expected utility amounts to a standard {\sl quadratic unconstrained binary optimization problem}~\cite{Lewis_QUBO_2022} of the real-symmetric matrix $C_{ij}(\vect{s})$. In general, this problem is strongly NP-hard~\cite{Lewis_QUBO_2022}. In the present case, we find that the matrix $C_{ij}(\vect{s})$ has, in the physically relevant regime, a large spectral gap and in practice admits a polynomial-time solution; see Appendix~\ref{app.QUBO}. Nevertheless, the problem requires computation of {\sl all} $C_{ij}(\vect{s})$ correlation functions and can be done, at reasonable computational cost, for lattice sizes up to $\d=20$ only. In the following, we introduce two practical, {\sl scalable} decoding schemes that approach an almost optimal decoder for lattice sizes up to $\d=256$ and $\d=64$, respectively.

\subsection{Minimum-energy heuristic}

On the Nishimori line, the Hamiltonian of Eq.~\eqref{eq:H_sc} reduces to Eq.~\eqref{eq.H_s}, i.e., $H(\vect{s},\vect{\tau})=H(\vect{s}\partial\vect{\tau})=H(\vect{c})$. Then, the energy of the fully polarized spin configuration ($\sigma_i=\pm1$ for all $i$) is proportional to the length of the correction chain, $H(\vect{c}) = -J\sum c_{ij}\propto -|\vect{c}|$. Hence, the MWPM can be interpreted as the decoder minimizing the energy of this spin configuration. 

To adapt this heuristic argument away from the Nishimori line, we note that the energy of the fully polarized spin configuration can be written as
\begin{align}\label{eq.H_sigma}
    \left. H(\vect{s},\vect{\tau})\right|_{\sigma_i\sigma_j=1} = - \sum_{\langle ij\rangle} c_{ij} w_{ij}\,,
\end{align} 
where $\vect{c}=\vect{s}\partial\vect{\tau}$ and we defined positive weights $w_{ij}=[\vect{s}J_{\vect{s}}]_{ij}$. To minimize the energy Eq.~\eqref{eq.H_sigma} with respect to $\vect{c}$, we deploy an MWPM algorithm on a {\sl weighted graph}, where each edge is assigned a weight that explicitly depends on the measured syndrome value
\begin{align}
    w_{ij}=\begin{cases}
        J_+,\quad &s_{ij}=1\\
        |J_-|,\quad &s_{ij}=-1
    \end{cases} \,.
\end{align} 
This syndrome-weighted MWPM (SW-MWPM) algorithm serves as a direct, physically motivated generalization of MWPM away from the Nishimori line. (Note that in the extreme case $t_A=t_B$, $J_+=\infty$, and the corrections only pass through antiferromagnetic bonds.) Crucially, SW-MWPM breaks the degeneracy between multiple shortest paths down to a much smaller set of lowest-{\sl weight} paths; see Fig.~\ref{fig:degeneracy} in Appendix~\ref{app.CNN}. Even for moderate lattice sizes, the degeneracy is frequently broken down to a single path,  whereas unweighted MWPM must arbitrarily choose one out of 10-100 degenerate shortest paths, often resulting in suboptimal macroscopic order.

\subsection{Neural-network decoder}

To resolve the issue of the remaining degeneracy---especially on the Nishimori line where $J_+=|J_-|$ and thus SW-MWPM is equivalent to unweighted MWPM---we deploy a convolutional neural network (CNN). Specifically, we devise a two-stage \cite{Meinerz2022} hybrid-CNN decoder (Fig.~\ref{fig:decoder}): First, the bare observed syndromes $\vect{s}$ are decoded using (SW-)MWPM to produce a baseline chain $\vect{c}$ and the correction $\vect{\tau}^{(1)}$ inside $\vect{s}\vect{c}$. Then, taking both the initial syndromes $\vect{s}$ and the baseline chain $\vect{c}$ as inputs, the CNN refines the correction to yield a higher-utility correction $\vect{\tau}^{(2)}$. Finally, the bit-flip correction is applied to the region $\vect{\tau}=\vect{\tau}^{(1)}\vect{\tau}^{(2)}$. The CNN is trained with the loss function $\mathcal{L}_M = -\langle \hat{M}^2 \rangle_{\vect{s},\vect{\tau}}$ and thus designed to maximize the utility. The specific network architecture and training details are given in Appendix~\ref{app.CNN}.

\section{Numerical simulations}
To compare the quality of the decoders, we rely on two primary benchmarks:  the decoding threshold and the value of the decoded magnetization squared.
The latter is an analytically tractable quantity that is closely related to the fidelity, which in the preparation of a GHZ state on a given device is the
operational benchmark one wants to maximize.
 
The threshold is an intrinsic property of a given decoder that defines the phase boundary beyond which the macroscopic order cannot be decoded. This boundary in the $(t_A,t_B)$ plane [Fig.~\ref{fig:setup}(b)] corresponds to the crossing point of the decoded magnetization squared lines for different system sizes; see Fig.~\ref{fig:decoded_M} below. To evaluate the threshold in a way that is less susceptible to finite-size effects, we retrieve it from a data collapse of the Binder cumulant of the magnetization~\footnote{The Binder cumulant is defined as $U=\frac{3}{2}\left(1-\frac{\langle \hat{M}^4\rangle}{3\langle \hat{M}^2\rangle^2}\right)$}. 

\begin{figure}[b]
    \centering
    \includegraphics{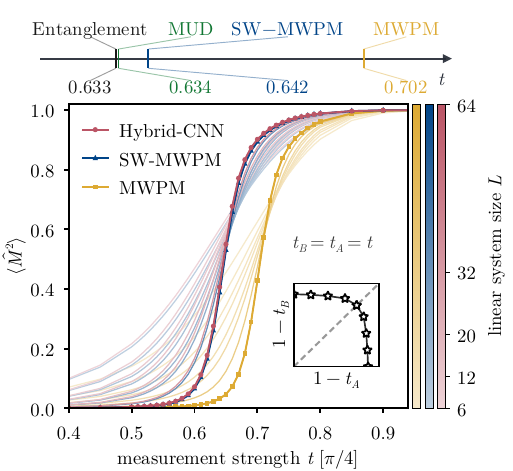}
    \caption{
    \textbf{Scalable decoding of long-range order.}
Decoded $\langle \hat{M}^2\rangle$ for $\d=6$--$64$ is plotted along the diagonal $t_A=t_B=t$. The strip on top marks the extrapolated thresholds $t_\text{c}$ of each decoder together with the decoder-independent {\sl entanglement} threshold (Appendix~\ref{app.dwe}). The SW-MWPM decoder nearly reaches the entanglement bound---recovering $87\%$ of the gap to bare MWPM [Eq.~\eqref{eq.threshold_gap}]---and together with the hybrid-CNN restores substantially more long-range order than bare MWPM up to $\d=64$, well beyond the reach of the optimal MUD.
    } 
    \label{fig:decoded_M}
\end{figure}

\subsection{Sampling the protocol}
Both the finite-size magnetization moments and the threshold extraction require simulation of the protocol over many shots and system sizes. Sampling the syndromes directly from $P(\vect{s})\propto Z_\text{RBIM}[\vect{s}]$ requires evaluation of the RBIM partition function and is prohibitively expensive. The joint distribution Eq.~\eqref{eq.P_s_sigma}, however, is easy to sample in the opposite order. First, we draw an unbiased spin configuration, $\sigma_i=\pm1$ with equal probability. Second, we draw each syndrome independently, conditioned on the relative alignment $\sigma_i\sigma_j$ of its two adjacent spins,
\begin{align}\label{eq.conditioned}
    P(s_{ij}\mid\sigma_i\sigma_j) =\frac{1-s_{ij}\cos 2(\sigma_i t_A + \sigma_j t_B)}{2} \,,
\end{align}
that is, $s_{ij}=1$ with probability $\sin^2(\sigma_i t_A+\sigma_j t_B)$. By construction, this reproduces Eq.~\eqref{eq.P_s_sigma} exactly, while avoiding the partition function altogether.

We pass the sampled syndromes $\vect{s}$ to the decoder, which returns a correction $\vect{\tau}$, and record the decoded magnetization of the shot, $M=\frac{1}{N}\sum_i \sigma_i\tau_i$. Averaging over shots yields $\langle \hat{M}^2\rangle$ and, analogously, the higher moment $\langle \hat{M}^4\rangle$ entering the Binder cumulant. With moderate computational resources we collect $N_\text{shots}=10^6$--$10^7$ configurations per point in the $(t_A,t_B)$ plane. The accessible system size is set by the decoders' complexity. For SW-MWPM, which is computationally equivalent to regular MWPM, we study systems with up to $256\times256$ data qubits. The subsequent CNN stage reduces accessible systems to $\d=64$ (Fig.~\ref{fig:decoded_M}). 
The exact MUD requires evaluating all-to-all spin correlations and is feasible for the linear system sizes of up to $\d=20$.

\subsection{Numerical results: Decodability threshold}
We extract the threshold by evaluating the decoded magnetization squared for multiple system sizes, from the smallest $L=4$ up to the largest system accessible to each decoder. In Fig.~\ref{fig:decoded_M}, we present a representative example of the decoded magnetization with the largest system size capped by the hybrid-CNN decoder.
For every decoder presented, we observe the expected behavior of perfectly recoverable long-range order, $\langle \hat{M}^2\rangle=1$, in the strong measurement limit $t\to\pi/4$, and an inability of the decoder to correct for weakly measured syndromes $t\to0$, where $\langle \hat{M}^2\rangle\to0$. 
Larger systems exhibit a more rapid transition, which for $\d\to\infty$ becomes a Heaviside theta function; while smaller systems have a more gradual change from ordered to disordered phase. The data for all of the systems cross at approximately one point that (roughly) corresponds to the threshold. 
To precisely extract the threshold for a given decoder, we perform a data collapse of the Binder cumulant of the magnetization; 
see Appendix~\ref{app.finite_size}.

Remarkably, simply weighting the syndromes in the SW-MWPM decoder yields a substantially lower (i.e., improved) threshold than conventional MWPM, as the crossing point shifts to weaker measurements (to the left). The subsequent CNN stage of the hybrid decoder results in a marginal additional improvement of the threshold. 

To map out an entire threshold phase diagram, we repeat the numerical experiment for fixed $t_B=0.7,0.8,0.9,1.0$ in units of $\pi/4$ and sweep $t_A$ to find the threshold $t^\text{c}_A(t_B)$ plotted in Fig.~\ref{fig:setup}(b). 
Similar to the diagonal line $t_A = t_B$ of Fig.~\ref{fig:decoded_M}, we find that SW-MWPM greatly improves the threshold that now closely follows the theoretical maximum inferred from the entanglement (decoder-independent) characteristics of the protocol; see Appendix~\ref{app.dwe}. We quantify the threshold improvement by comparing how much closer the threshold of the SW-MWPM decoder is to the optimal threshold than the one of the conventional MWPM. We compute this relative `threshold gap' as
\begin{align}\label{eq.threshold_gap}
    {\cal T} = \frac{t^\text{c}_\text{SW-MWPM} - t^\text{c}_\text{MWPM}}{t^\text{c}_\text{opt}-t^\text{c}_\text{MWPM}}
\end{align}
and find that for $t_B=0.7,0.8,0.9,1.0$ in units of $\pi/4$, the relative threshold gaps are ${\cal T} = 86\pm 1\%,85\pm 1\%, 64\pm 6 \%, 0 \%$, respectively. 
The largest value is found on the diagonal $t_A=t_B$, where ${\cal T} = 87\pm1\%$ indicates that about 7/8 of the gap between the MWPM-threshold and the optimal threshold can be overcome by the SW-MWPM decoder.

Finally, we observe that the MWPM phase boundary is roughly circular, i.e.\ equidistant from the strong measurement limit, $t_A=t_B=\pi/4$, at the origin of Fig.~\ref{fig:setup}(b). This property follows from the gauge `obliviousness' of the MWPM decoder that discards the precise syndrome configuration and only uses the gauge-invariant flux. In Appendix~\ref{app.mwpm}, we show that the phase boundary of any gauge-oblivious decoder follows a curve $\sin 2t_A\sin 2t_B=\text{const.}$ (with $\text{const.}=0.794$ for the MWPM) in the leading order in a small parameter $[\cos 2t_A\cos 2t_B]^4\leq 1.8\times 10^{-3}$ in the physically relevant regime.

\subsection{Numerical results: Decoded magnetization squared}
In a practical context, a more important figure of merit than the threshold might be the average decoded magnetization squared (which is connected to the fidelity of the final state). To quantify the performance of a  new decoder (dec-1) relative to the baseline decoder (dec-2), we define the relative improvement 
\begin{align}\label{eq.delta_rel}
    \Delta_\text{dec-1, dec-2} = \frac{\langle \hat{M}^2\rangle_\text{dec-1}-\langle \hat{M}^2\rangle_\text{dec-2}}{\langle \hat{M}^2\rangle_\text{dec-1}+\langle \hat{M}^2\rangle_\text{dec-2}} \,.
\end{align}
When two decoders have different thresholds $t^{1/2}_c$, there is an intermediate regime $t^1_c<t<t^2_c$ where the system size affects the decoded order parameter oppositely. In the extreme case, $\langle \hat{M}^2\rangle_\text{dec-1}=1$ and $\langle \hat{M}^2\rangle_\text{dec-2}=0$ (which is realized for $\d\to\infty$), the relative improvement reaches $\Delta_\text{dec-1, dec-2}=100\%$.

\begin{figure}
    \centering
    \includegraphics{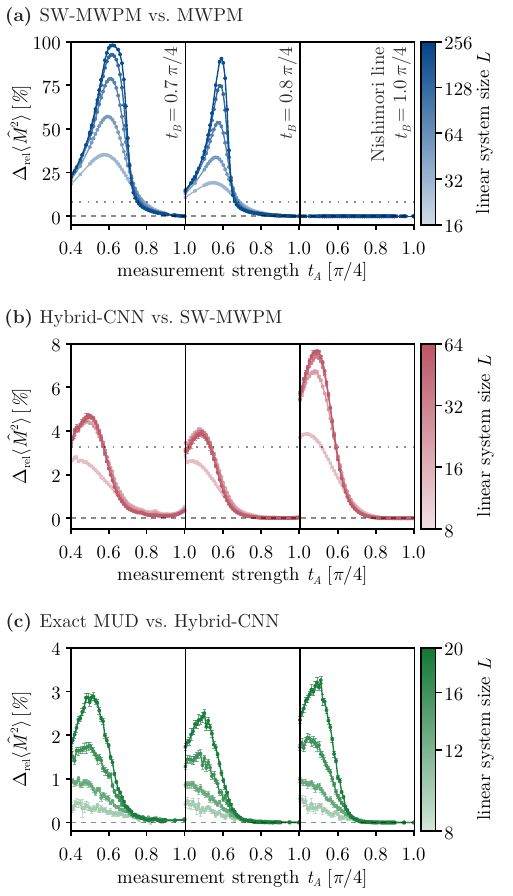}
    \caption{
    \textbf{Relative decoder performances.} Shown is the normalized improvement in decoded squared magnetization \mbox{$\Delta_{\mathrm{rel}}\langle \hat{M}^2\rangle \equiv \Delta_{\mathrm{dec-1,dec-2}}$} defined in Eq.~\eqref{eq.delta_rel}. (a) Syndrome-weighted MWPM relative to conventional MWPM. (b) Hybrid-CNN decoder relative to syndrome-weighted MWPM. (c) Exact MUD relative to the Hybrid-CNN decoder. Each column corresponds to a cut through the phase diagram of Fig.~\ref{fig:setup}(b), with varying measurement strength $t_A$ at fixed $t_B$. The color bars indicate the lattice sizes.
    }
    \label{fig:decoders_improvement}
\end{figure}

In Fig.~\ref{fig:decoders_improvement}, we compare the decoders in the order of increasing numerical complexity, and also increasing accuracy. In the top panel, we compare the syndrome-weighted and the standard MWPM decoders. On the Nishimori line (rightmost panel), the two decoders are equivalent since weights $|J_+|=|J_-|$. Going away from the Nishimori line, the relative improvement reaches almost maximal $\Delta_\text{rel}\langle \hat{M}^2\rangle  = 100\%$ for the largest systems studied in the intermediate regime $t^{\text{SW-MWPM}}_A~<~t_A~<~t^{\text{MWPM}}_A$. 

In the middle panel, we quantify the CNN performance in the hybrid decoder by subtracting the magnetization squared decoded with only the first stage, versus the  SW-MWPM decoder as baseline. The largest improvement of $\Delta_\text{rel}\langle \hat{M}^2\rangle~\approx~8\%$ is observed on the Nishimori line, where syndrome weights are trivial. Away from the Nishimori line, the improvement decreases; nevertheless, it is not negligible. These benefits, however, increase the numerical complexity of the decoder and limit the system sizes to $L\leq 64$. 

Finally, in panel (c) of  Fig.~\ref{fig:decoders_improvement}, we analyze how close the hybrid-CNN decoder approaches the theoretically possible maximum given by the optimal MUD decoder. The improvement for all the systems and regimes is under $\Delta_\text{rel}\langle \hat{M}^2\rangle~\lesssim~3\%$, while the decoding complexity is significantly larger and does not allow us to study large systems $\d > 20$.

\section{Discussion}
\label{sec:discussion}

Our results recast decoding as a Bayesian decision problem in which the decoder is fixed by the utility function encoding the operational goal. Viewed this way, MUD is not specific to GHZ-state preparation: any task whose success admits a {\sl quantitative} figure of merit can be decoded optimally by maximizing the corresponding expected utility.

This directly connects to conventional {\sl quantum error correction}. For the 2D repetition code, we show that MUD reproduces the standard decoders as special cases; see Appendix~\ref{app.repetition}. The utility $\mathcal{U}_{\mathrm{MLD}}(\bm{\tau},\bm{\sigma}) = \theta\!\left(\sum_i \sigma_i \tau_i\right)$ recovers maximum-likelihood decoding, whereas the simpler magnetization utility $\mathcal{U}_{M}(\bm{\tau},\bm{\sigma}) = \frac{1}{N}\sum_i \sigma_i \tau_i$ yields a per-qubit decoder with optimum $\tau_i = \operatorname{sign}\langle \sigma_i \rangle_{\bm{s}}$---the blockwise and bitwise maximum a posteriori estimators, respectively. Because the bitwise decoder avoids the nonlinearity of the $\theta$-function, we expect it to deliver competitive logical fidelity at a fraction of the cost; a quantitative investigation of MUD applied to the 2D repetition code is left to future work.

A second consequence is {\sl hardware-aware} decoding. On NISQ hardware, the feedforward correction is applied with imperfect gates, so a recovery that is optimal under ideal application may cease to be optimal once the errors introduced by its own execution are taken into account. Whereas standard decoders use device-characterization data to build a noise model, MUD can leverage the same data one step further, encoding the faults of the recovery operation itself into the utility function and returning the correction that is optimal for the actual noisy hardware.

A further direction opens up when decoding logical operations rather than a static memory. For a logical gate acting jointly on several code blocks---as in transversal entangling gates~\cite{Bryan_Transversal_2009,Bluvstein_Logical_2024} or lattice surgery~\cite{Horsman_Surface_2012,Litinski_game_2019}---the fidelity of the executed gate is a natural utility. The MUD then optimizes the success of the operation itself rather than the recovery of each logical qubit in isolation. By promoting the objective of decoding to an explicit design choice, MUD provides a template for such operation-tailored decoders.\\

\textit{Data availability.---}
The numerical data shown in the figures are available on Zenodo~\cite{zenodo_MUD}. \\

\textit{Acknowledgments.---}
We thank Guo-Yi Zhu for discussions.
M.Y. acknowledges support from an ML4Q postdoctoral fellowship.
The Cologne research group is supported, in part, by
the Deutsche Forschungsgemeinschaft (DFG, German Research Foundation) under Germany’s Excellence Strategy—Cluster of Excellence Matter and Light for Quantum Computing (ML4Q) EXC 2004/1 -- 390534769, the CRC network TR 183 (Project Grant No.\ 277101999), 
as well as Research Unit FOR5919 (Project Grant No.\ 550495627). 
Our numerical simulations were performed on the RAMSES cluster at RRZK Cologne. 

\appendix

\section{Maximum-utility decoder}

In this appendix, we provide additional details on the maximum-utility decoder (MUD) and derive certain properties cited in the main text. 

\subsection{The expected utility as a quantum expectation value}
\label{app.MUD} 

We first explicitly derive the relation between the quantum expectation value of the magnetization squared, Eq.~\eqref{eq:M2_sc}, and the expected utility in Eq.~\eqref{eq:utility_function}. Using the gauge symmetry, the partition function $\tilde{Z}[\vect{s},\vect{\tau}]$ of the Hamiltonian Eq.~\eqref{eq:H_sc} can be explicitly rewritten as
\begin{align}
    \tilde{Z}[\vect{s},\vect{\tau}] 
    &= \sum_{\vect{\sigma}}e^{\beta \sum_{\langle ij \rangle} J_{s_{ij}}\tau_{i}\sigma_i\tau_{j}\sigma_j} 
    \\&= \sum_{\vect{\sigma}'}e^{\beta \sum_{\langle ij \rangle} J_{s_{ij}}\sigma'_i\sigma'_j} = Z_\text{RBIM}[\vect{s}],
\end{align}
where $Z_\text{RBIM}[\vect{s}]$ is a standard RBIM partition function with Eq.~\eqref{eq.H_s}, and we changed variables as $\sigma_i'=\sigma_i\tau_i$. Then, we express the conditional probability in terms of this partition function
\begin{align}
    P(\vect{\sigma}|\vect{s}) \equiv \frac{P(\vect{\sigma}, \vect{s})}{\sum_{\vect{\sigma}}P(\vect{\sigma}, \vect{s})}
    =
    \frac{e^{\beta\sum_{\langle ij \rangle} J_{s_{ij}}\sigma_i\sigma_j}}{Z_\text{RBIM}[\vect{s}]}.
\end{align}
Finally, for the expected utility, we obtain
\begin{align}
    &\mathbb{E}[{\cal U}] \equiv \sum_{\vect{\sigma}} \mathcal{U}(\vect{\tau},\vect{\sigma}) P(\vect{\sigma}|\vect{s})
    \\&=\sum_{\vect{\sigma}} \left[\frac{1}{N}\sum_i\sigma_i \tau_i\right]^2  \frac{e^{\beta\sum_{\langle ij \rangle} J_{s_{ij}}\sigma_i\sigma_j}}{Z_\text{RBIM}[\vect{s}]}\label{eq:A6}
    \\&=\frac{\sum_{\vect{\sigma}'} \left[\frac{1}{N}\sum_i\sigma'_i \right]^2  e^{\beta\sum_{\langle ij \rangle} J_{s_{ij}}\tau_i\tau_j\sigma'_i\sigma'_j}}{Z_\text{RBIM}[\vect{s}]}\equiv \langle \hat{M}^2 \rangle_{\vect{s},\vect{\tau}}.
\end{align}
This equivalence proves that maximizing the long-range order of the decoded state is achieved via a decision based on the maximum expected utility. 

\subsection{Exact MUD}\label{app.QUBO} 

The equivalence of quantum and classical expectation values allows us to reformulate the exact MUD decision in terms of a standard computational problem---the quadratic unconstrained binary optimization~\cite{Lewis_QUBO_2022}, which in general is strongly NP-hard. The expected utility in the form Eq.~\eqref{eq:A6} can be expressed in terms of the spin-spin correlation functions as
\begin{align}
    \mathbb{E}[{\cal U}] = \sum_{\vect{\sigma}} \left[\frac{1}{N}\sum_i\sigma_i \tau_i\right]^2  \frac{e^{\beta\sum_{\langle ij \rangle} J_{s_{ij}}\sigma_i\sigma_j}}{Z_\text{RBIM}[\vect{s}]}
    \\=
    \frac{1}{N^2}\sum_{i,j}\tau_i \tau_j \frac{\sum_{\vect{\sigma}} \sigma_i\sigma_j  e^{\beta\sum_{\langle ij \rangle} J_{s_{ij}}\sigma_i\sigma_j}}{ Z_\text{RBIM}[\vect{s}]}
    \\=
    \frac{1}{N^2}\sum_{i,j}\tau_i C_{ij}(\vect{s})  \tau_j \equiv \frac{1}{N^2} M_{\hat{C}}(\vect{\tau})
\end{align}
where $C_{ij}(\vect{s}) = \langle\sigma_i \sigma_j\rangle_{\vect{s}}$, which we compute via the Kac-Ward formula~\cite{
Kasteleyn_dimer_1963,Loh_Efficient_2006,Thomas_Exact_2009,Kager_Loop_2013,ChelkakC_Revisiting_2017}. Then, the optimal correction $\tau_i =\pm 1$ maximizes the binary quadratic form $M_{\hat{C}}$ of the real-symmetric matrix $C_{ij}(\vect{s})$.

\begin{figure}[t!]
    \centering
    \includegraphics{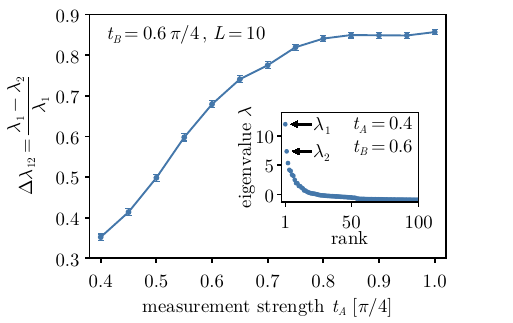}
    \caption{{\bf Correlation-matrix spectral gap.} The relative gap $\Delta\lambda_{12}=(\lambda_1-\lambda_2)/\lambda_1$ between the two largest eigenvalues of the correlation matrix $C_{ij}^{(s)}=\langle\sigma_i\sigma_j\rangle_s$ is plotted as a function of $t_A$, for fixed $t_B=0.6\,\pi/4$ and lattice size $\d=10$. The data $\Delta\lambda_{12}$ are averaged over 500 disorder configurations, with standard-deviation error bars. The inset contains the ranked eigenvalue spectrum of $C_{ij}^{(s)}$ for a representative measurement configuration at $t_A=0.4\,\pi/4$ and $t_B=0.6\,\pi/4$. }
    \label{fig:spectrum}
\end{figure}

For an arbitrary matrix $C_{ij}(\vect{s})$, the problem is intractable even for moderate linear system sizes, since the number of corrections $\{\tau_i\}$  is $2^{\d^2}$. However, in the ordered phase, the matrix $\hat{C}$ is rank-1 dominant and has a large spectral gap; see Fig.~\ref{fig:spectrum}. Moreover, we empirically find that this holds even in the disordered phase not too far from the transition. In this case, we can typically find the solution in polynomial time following the program described below.    

\textit{Local iterative optimization.---}
The spectrum of $\hat{C}$ is dominated by a single large eigenvalue; see Fig.~\ref{fig:spectrum}. Therefore, the eigenvector $\vect{v}$ with the largest eigenvalue provides a good starting point $\vect{\tau}^{(0)} = \operatorname{sign}(\vect{v})$ for an iterative optimization procedure. We iteratively flip a random spin $\tau_i$ if $\Delta M_i \equiv -4\tau_i \left(\sum_{j\neq i} C_{ij}\tau_j\right)> 0$ until no spin can be flipped. We restart the optimization process multiple times and select the best-performing solution $\vect{\tau}^*$. 

\textit{Certifying the solution.---}
The output of the optimization procedure $\vect{\tau}^*$ is not proven to be optimal, because there may be a global maximum that local updates do not reach even after multiple restarts. However, the solution $\vect{\tau}^*$ can be easily certified, i.e., proven to be optimal if it is a global maximum. To this end, we define a vector $d_i=\tau^*_i[\hat{C}\vect{\tau}^*]_i$ and a matrix $S_{ij} = \delta_{ij} d_i -C_{ij}$. If the matrix $\hat{S}$ is positive semidefinite, then the candidate $\vect{\tau}^*$ is a global maximum of the quadratic form $M(\vect{\tau}^*)$. 

We now show that for a positive-semidefinite $\hat{S}$, the quadratic form is bounded by $M(\vect{\tau}^*)$, i.e., $\vect{\tau}^*$ is a global maximum. In other words, for any candidate solution $\tau_i=\pm 1$, we show that $M(\vect{\tau})\leq M(\vect{\tau}^*)$. Expressing $\hat{C} = \operatorname{diag}(\vect{d})-\hat{S}$, we find that
\begin{align}
\vect{\tau}^T\hat{C}\vect{\tau} = 
\vect{\tau}^T\operatorname{diag}(\vect{d})\vect{\tau}-\vect{\tau}^T\hat{S}\vect{\tau}
\leq \sum_{i}d_i\equiv M(\vect{\tau}^*)
\end{align}
for positive-semidefinite $\hat{S}$, which satisfies $\vect{\tau}^T\hat{S}\vect{\tau}\geq0$. Here, we used 
\begin{align}
    \vect{\tau}^T\operatorname{diag}(\vect{d})\vect{\tau} = \sum_{i,j} \tau_i \delta_{ij}d_i\tau_j=\sum_{i} \tau_i^2 d_i=\sum_{i}d_i,
\end{align}
for any $\tau_i=\pm1$, and by definition 
\begin{align}
    \sum_{i}d_i = \sum_{i,j}\tau^*_iC_{ij}\tau^*_j\equiv M(\tau^*).
\end{align}

Note that positive semidefiniteness of $\hat{S}$ proves that $\vect{\tau}^*$ is optimal; however, the converse is not true, and an indefinite $\hat{S}$ does not prove that $\vect{\tau}^*$ is not optimal. 
 
\textit{Branch and bound.---}
When the certificate is inconclusive, we prove or improve $\vect{\tau}^*$ with an exact branch-and-bound search~\cite{Land_Discrete_1960, Krislock_BiqCrunch_2017}. Since $\vect{\tau}^*$ is almost always already optimal and $\hat{C}$ is rank-1 dominant, the search tree typically collapses, and the proof costs far less than assembling the correlation matrix $\hat{C}$. If the search exceeds a
predefined node budget, the algorithm returns $\vect{\tau}^*$ without a
proof of optimality; the returned correction remains valid and at least of
the heuristic quality.

\textit{Practice.---}
In the regime of sufficiently strong measurements where the GHZ state can be recovered, the local iterative procedure almost always finds the optimal solution. The certificate and branch-and-bound stages are primarily used to prove that the solution is optimal. 

To test this, we perform a benchmark simulation of the algorithm. We randomly choose $10^4$, $10^3$, and $10^2$ points $t_A,t_B \in [0.4\,\pi/4, \pi/4]$ for $L=8,16$ and $20$, respectively. For each point and system size, we draw an uncorrelated sample from the distribution Eq.~\eqref{eq.P_s_sigma}. We then decode the obtained syndromes with the optimal MUD algorithm and collect the statistics. We find that the certification fails in $19.8\%,28.7\%$, and $27.0\%$ of the samples, respectively. Nevertheless, the local iterative solution $\vect{\tau}^*$ was proven to be a global maximum in $99.5\%,82.9\%$, and $79\%$ of cases, respectively. For all three systems, we find that all of the cases where the iterative solution is either not optimal or where the branch-and-bound exceeds the node budget occur below or near the threshold, i.e., in the paramagnetic phase. We expect that in this regime, the search terminates without improving $\vect{\tau^*}$ due to the presence of near-degenerate optima; thus, the deviation from the true optimum is negligible.  

\textit{Correlation-function truncation.---}
The runtime of the decoder is dominated by computing the all-to-all correlation-function matrix $C_{ij}$. This cost can be greatly reduced by truncating the matrix $\hat{C}$ by setting $C_{ij}=0$ for $|i-j|>r_\text{max}$ without computing the correlation function. For $r_\text{max}=4$ we find that the algorithm produces comparable results with a much shorter runtime for large systems; see Fig.~\ref{fig:decoding_extrapolation}. This approximation is justified as away from the strong measurement limit, the spin-spin correlation function decays with the distance between spins. 

\subsection{Performance of the optimal and truncated MUD decoders}
\begin{figure}[t]
    \centering
    \includegraphics{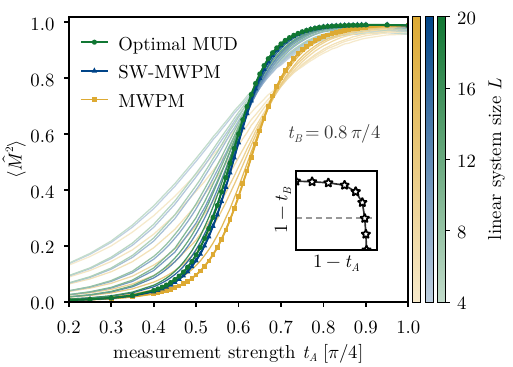}
    \caption{\textbf{Benchmarking the scalable decoders against the optimal MUD.} Average decoded magnetization squared $\langle \hat{M}^2\rangle$ is shown as a function of the measurement strength $t_A$ at fixed $t_B=0.8\,\pi/4$. Data are shown for the exact optimal MUD (green), the syndrome-weighted MWPM (SW-MWPM, blue), and bare MWPM (yellow); within each decoder, the color shading encodes the linear system size $\d=4$--$20$. The scalable SW-MWPM closely tracks the optimal MUD across the entire cut, and both restore markedly more long-range order than bare MWPM.}
    \label{fig:decoder_exact}
\end{figure}

The exact optimal and truncated MUD decoders are much more numerically demanding than the SW-MWPM or even hybrid-CNN decoders. The former two we study for the linear system sizes of up to $\d=20$ and $\d=32$, respectively, while the latter two are amenable to the lattice sizes $\d=256$ and $\d=64$, respectively. The optimal MUD yields the maximal possible per-shot decoded magnetization squared by explicitly maximizing the quadratic form of the correlation matrix; see Fig.~\ref{fig:decoder_exact}. Nevertheless, the SW-MWPM and hybrid-CNN perform almost identically to the optimal MUD, with the deviations Eq.~\eqref{eq.delta_rel} between them being $\Delta_\text{rel}\lesssim 10\%$ and $\Delta_\text{rel}\lesssim 3\%$, respectively; see Fig.~\ref{fig:decoders_improvement}. At the same time, the thresholds of the SW-MWPM and the optimal and truncated MUD closely follow each other when accounting for finite-size effects; see Fig.~\ref{fig:decoding_extrapolation}.

\section{Hybrid-CNN decoder}
\label{app.CNN}

In this appendix, we provide additional details on the hybrid-CNN decoder as a practical implementation of the ersatz MUD decoder. To make a scalable, nearly optimal decoder, we use the approach that leverages neural networks for decoding. We first note that the joint probability Eq.~\eqref{eq.P_s_sigma} yields a non-trivial dependence on  $\vect{s}$ of the conditional distribution
\begin{align}
    P(\vect{\sigma}|\vect{s}) = \frac{P(\vect{\sigma}, \vect{s})}{\sum_{\vect{\sigma}}P(\vect{\sigma}, \vect{s})}.
\end{align}
This implies that $\vect{s}$ contains a learnable signal for predicting $\vect{\sigma}$. (Note that the distribution of the gauge-invariant vortices $\vect{f}=\partial s$ does not contain any information about $\vect{\sigma}$ since $P(\vect{\sigma}|\vect{f})=\frac{1}{2^N}$.) The problem possesses translational invariance (up to the boundary of the sample); hence, we use a convolutional neural network (CNN) architecture. 

\begin{figure}[t]
    \centering
    \includegraphics{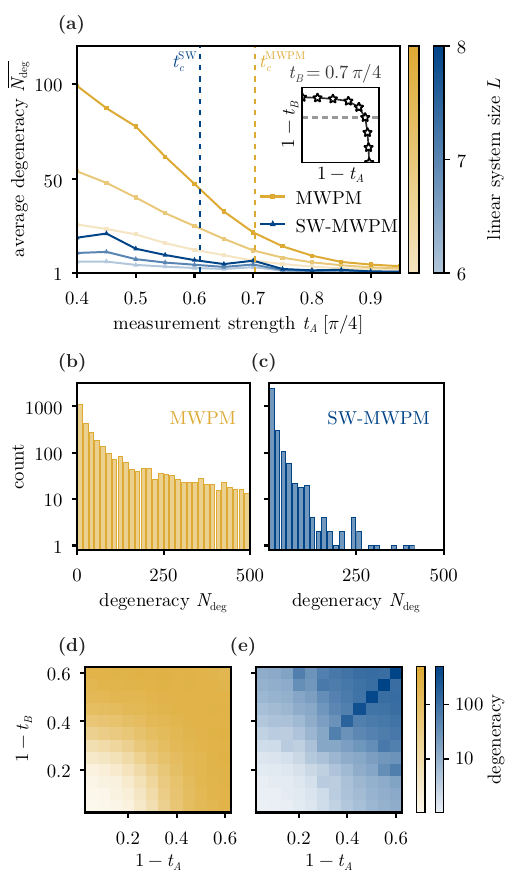}
    \caption{\textbf{Degeneracy of minimum-weight corrections.} (a) The average degeneracy $\overline{N_{\mathrm{deg}}}$ of minimum-weight correction chains is plotted as a function of $t_A$ at fixed $t_B=0.7 \pi/4$ for MWPM and syndrome-weighted MWPM. Color shades indicate the lattice size $\d$, and dashed vertical lines mark the corresponding threshold estimates. The inset indicates the horizontal cut through the phase diagram at $t_B=0.7$. Panels (b) and (c) show the distribution of degeneracies over individual disorder snapshots for $\d=8$ and $t_A=0.5$. Panels (d) and (e) show the average degeneracy in the $(t_A,t_B)$ plane for $\d=10$ for MWPM and syndrome-weighted MWPM, respectively, with color shade encoding $\overline{N_{\mathrm{deg}}}$.
    }
    \label{fig:degeneracy}
\end{figure}

A practical challenge in training the model is the presence of a large number of antiferromagnetic bonds. (Recall that Eq.~\eqref{eq.P_s_sigma} implies that any given syndrome has a marginal probability $P(s_{ij})=\frac{1+s_{ij}\cos2t_A\cos2t_B}{2}$, which is close to $1/2$ near $t_A=t_B=\pi/4$). However, most antiferromagnetic bonds belong to closed domain walls that can be trivially eliminated by flipping all the spins inside them. Hence, to simplify training, we first apply the SW-MWPM decoder, which identifies all closed antiferromagnetic domains with $\vect{\tau}^{(1)}$. The deformed chain $\vect{c}=\vect{s}\partial\vect{\tau}^{(1)}$ now contains only antiferromagnetic bonds that belong to open chains spanning between the fluxes. To focus the model on these non-trivial chains, we pass the original syndromes $\vect{s}$ and the smallest-weight deformed chain $\vect{c}$ to the CNN that outputs predictions for the correction $\vect{\tau}$ that maximizes $\langle \hat{M}^2 \rangle$.

Note that $\vect{c}$ is chosen arbitrarily by the algorithm from a large pool of chains with the same (weight) length; see Fig.~\ref{fig:degeneracy}. Although reducing the length of the domain walls increases the magnetization squared, the domain walls of the same length can yield different order parameters. The objective of the CNN is to resolve this ambiguity in $\vect{c}$ (and thus in $\vect{\tau}^{(1)}$, the boundary of $\vect{s}\vect{c}=\partial\vect{\tau}^{(1)}$) by calibrating the choice with $\vect{\tau}^{(2)}$ to obtain the largest expected magnetization squared.

\subsection{Model and training details}

\textit{Model.---}
Specifically, we use the residual CNN with Feature-Wise Linear Modulation (FiLM) to encode the information of the physical parameters $t_A$ and $t_B$. As a compromise between efficiency and runtime, we chose a $3\times3$ kernel with 12 blocks and 64 channels (normalized over 8 groups). The total number of learnable parameters is 990,401. We find that the number of parameters needed is significantly larger than typically required for toric code decoding, and reducing the number deteriorates the performance of our architecture. 

\textit{Features.---}
To embed $2\d(\d-1)$ edges into a $\d\times \d$ square lattice, we separate edges into vertical and horizontal, constituting two $\d\times(\d-1)$ channels. Further, we map the edge $\d\times(\d-1)$ to the vertex $\d\times \d$ configurations by assigning to each vertex in horizontal/vertical channels the sum of adjacent horizontal/vertical edges. The resulting input of the model is zero-padded for $\d\times \d$ channels for $\vect{s}$ and $\vect{c}$ edge configurations in addition to $t_A,t_B$ passed through the FiLM layers.

\begin{figure}[t]
    \centering
    \includegraphics{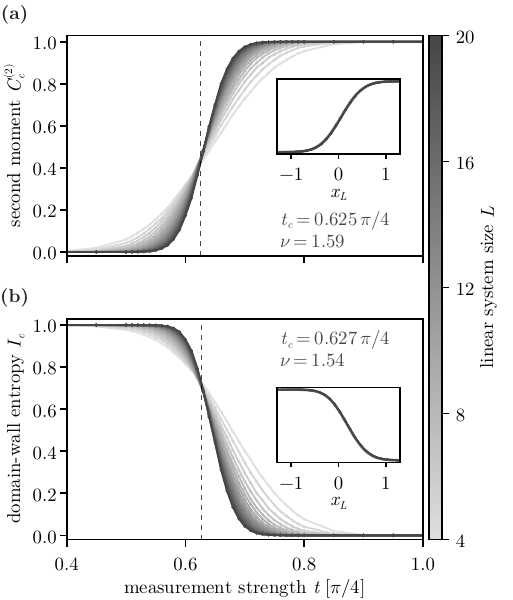}
    \caption{{\bf Second moment $C^{(2)}_\text{c}$ and domain-wall entropy $I_\text{c}$.} The data are presented along the radial line $t_A=t_B=\pi/4-t_{\mathrm{rad}}$ . The upper plot (a) depicts the second moment of $C^{(2)}_\text{c}$ plotted against the measurement strength $t_{\mathrm{rad}}$ for different linear system sizes $\d$. Panel (b) shows the domain-wall entropy for the same cut. The color shades encode the lattice size, and the insets show the corresponding finite-size scaling collapse as a function of the rescaled variable $x_{\d} = (t_{\mathrm{rad}}-t_\text{c})\d^{1/\nu}/t_\text{c}$.
}
    \label{fig:domain_wall_entropy}
\end{figure}

 \begin{figure}[t]
    \centering
    \includegraphics{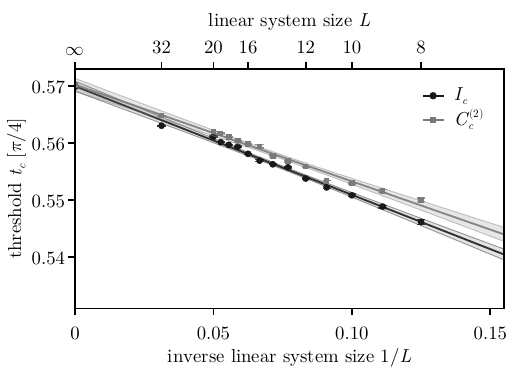}
    \caption{{\bf Finite-size analysis of the entanglement transition.} 
    Each point represents an estimate of the $t^\text{c}_A$ threshold at fixed $t_B=0.9 \pi/4$ for a lattice size $\d$. Error bars on the data indicate the bootstrap uncertainty of the threshold. The solid lines show linear fits in the inverse linear size of the lattice $1/\d$, and shaded bands represent the uncertainty of the fit. 
    When extrapolated to the thermodynamic limit $\d \to \infty$, both observables converge to the same value within numerical uncertainty. The averaged value of the threshold obtained from the two observables is used in the main text. 
   }
    \label{fig:finite_size}
\end{figure}

\textit{Loss function.---}
For each shot of the experiment $(s_{ij},\sigma_i)$, the objective of the model is to predict the correction $\tau^{(2)}_i$ that would result in the largest {\sl averaged} decoded magnetization squared
\begin{align}
    -\mathcal{L}_M = \left(\sum_i \tau^{(1)}_i\tau^{(2)}_i\sigma_i\right)^2,
\end{align}
where $\tau^{(1)}_i$ is the first-stage corrections inside the loops $\vect{s}\vect{c}$. Perfectly predicting $\tau^{(2)}_i=\tau^{(1)}_i\sigma_i$ in each shot is not mathematically possible since away from the strong measurement point $t_A=t_B=\pi/4$, the probability distribution $P(\vect{\sigma}|\vect{s})$ is {\sl not} a delta function $\delta(\sigma_i\sigma_j - s_{ij})$. Hence, the ultimate goal of training is to minimize the loss function $\mathcal{L}_M$ averaged over a sufficient number of shots.

To make training more efficient in the initial stages, we introduce additional terms
\begin{align}
    \mathcal{L}_\text{BCE} = \operatorname{BCE}(\tau^{(2)}_i,\sigma_i\tau^{(1)}_i),\\
    \mathcal{L}_\text{align} = -\sum_{\langle ij \rangle} \tau^{(2)}_i \tau^{(2)}_j [c]_{ij}.
\end{align}
The binary cross-entropy term $\mathcal{L}_\text{BCE}$ produces a large number of gradients that facilitate fast learning. The edge-alignment term $\mathcal{L}_\text{align}$ penalizes the model for flipping domains that are not aligned with the remaining antiferromagnetic frustrations $c$. This term ensures that the model does not introduce new domains away from the frustrations.

We train the model by scheduling the loss function by smoothly changing the weights of the loss functions starting from $(\lambda_M,\lambda_\text{BCE},\lambda_\text{align}) = (0.3,0.3,0.2)$ at the first epoch to $(1.0,0.02,0.04)$ at the last epoch.

\begin{figure*}[t]
    \centering
    \includegraphics{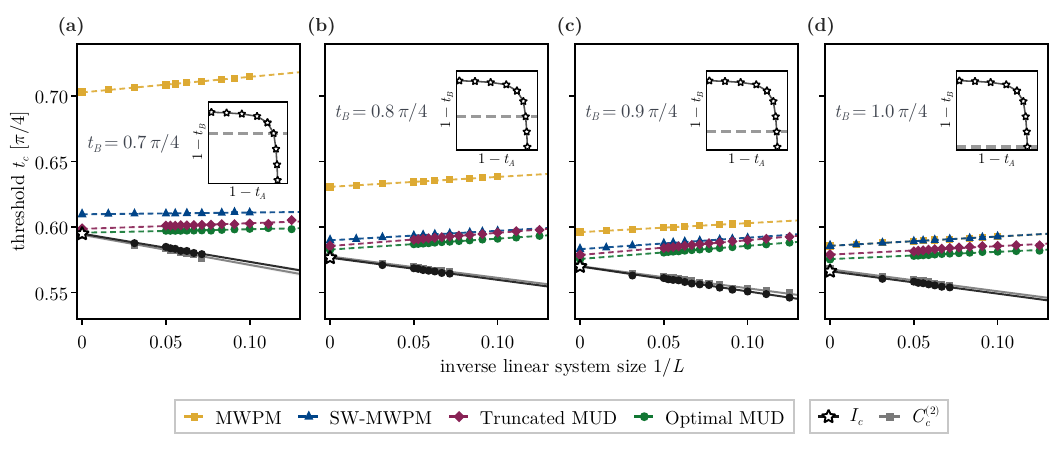}
    \caption{\textbf{Finite-size analysis of thresholds.} The finite-size thresholds obtained from the expanding window technique are plotted. The entanglement thresholds are computed from the data collapse of the domain-wall entropy $I_c$ and boundary-boundary correlation functions $C^{(2)}_c$. The decoding thresholds are extrapolated for the MWPM, SW-MWPM, the exact optimal MUD decoder, and the $r_\text{max}=4$ truncated MUD decoder. On the Nishimori line ($t_B=\pi/4$), panel (d), all thresholds are relatively close to each other. The performance of the bare MWPM decoder significantly degrades away from the Nishimori line, while the SW-MWPM, optimal MUD, and truncated MUD decoders maintain close thresholds to the entanglement transition. 
    }
    \label{fig:decoding_extrapolation}
\end{figure*}

\section{Entanglement threshold}
\label{app.dwe}

We compute the optimal threshold as the location of the {\sl entanglement transition} using the domain-wall entropy~\cite{Putz_Learning_2026}, which is equivalent to the coherent mutual information. We augment the square $\d\times \d$ lattice by coupling the left (right) boundary spins $\sigma_{x,1}$ ($\sigma_{x,\d}$) to an additional spin $\sigma_L$ ($\sigma_R$); see Fig.~\ref{fig:domain_wall_entropy}. The couplings between these new spins and the boundaries are the same as for the bulk spins. For a fixed syndrome configuration $\vect{s}$, we compute the correlation function $C_{\vect{s}}=\langle\sigma_L \sigma_R\rangle_{\vect{s}}$. The domain-wall entropy of this syndrome configuration is computed as 
\begin{align}
    I_{\vect{s}} = -\frac{1+C_{\vect{s}}}{2}\log_2 \frac{1+C_{\vect{s}}}{2}-\frac{1-C_{\vect{s}}}{2}\log_2 \frac{1-C_{\vect{s}}}{2}. 
\end{align}
We then average the domain-wall entropy and squared correlation function over the syndrome configurations:
\begin{align}
    I_c = \sum_{\vect{s}} I_{\vect{s}} P(\vect{s}), \qquad
    C^{(2)}_c = \sum_{\vect{s}} [C_{\vect{s}}]^2 P(\vect{s}).
\end{align}

\subsection{Numerical technique}

Since we are interested in the nonlinear moments of $C_{\vect{s}}$, we cannot use the procedure defined in the main text, which relies on computing averages of the joint distribution $P(\vect{\sigma},\vect{s})$. Therefore, for a specific configuration $\vect{s}$, we compute $C_{\vect{s}}$ as a contraction of the tensor network closely following Ref.~\cite{Zhu_Nishimori_2023}. We define the tensor network on a $(\d+2)\times \d$ lattice. We then force the left and right boundary spins to be aligned by inserting the identity bond matrix corresponding to an infinite ferromagnetic coupling strength. This ensures that $\sigma_{0,y}\equiv\sigma_L$ and $\sigma_{\d+1,y}\equiv\sigma_R$ for all $y$. Contracting the tensor network on this augmented lattice results precisely in $C_{\vect{s}}$.

The efficient numerical sampling procedure does not allow us to compute the nonlinear expectation value. However, it is well suited to sampling independent syndrome configurations $\vect{s}$ from $P(\vect{s})$, where we simply discard the latent variable $\sigma_i$ used for conditioning in Eq.~\eqref{eq.conditioned}. (To obey the boundary conditions at $x=0$ and $x=\d+1$, we sample only one spin, $\sigma_L$ and $\sigma_R$, per boundary.)  We compute $I_c$ and  $C^{(2)}_c$ by averaging over $N_\text{shots}=50,000$ independent syndrome configurations for  $\d=4$--$20$.

\subsection{Finite-size analysis}\label{app.sub.finite_size}

Both $I_c$ and  $C^{(2)}_c$ exhibit threshold behavior shown in Fig.~\ref{fig:domain_wall_entropy}. However, even on the Nishimori line, the thresholds obtained from the data collapse for $\d\leq20$ deviate from the value reported in the literature. We attribute this discrepancy to finite-size effects and open boundary contributions. 

We extrapolate the data to the thermodynamic limit to take into account these corrections. We perform the data collapse over the expanding window of data using system sizes $\d=4$~--~$\d_0$ with $\d_0=7$--$20$. Then, we plot the obtained threshold values $t_A^\text{c}(\d_0)$ versus $\frac{1}{\d_0}$ for both $I_c$ and  $C^{(2)}_c$; see Fig.~\ref{fig:finite_size} for $t_B=0.9\pi/4$. Although the finite-size thresholds obtained from the two observables differ significantly, their thermodynamic extrapolations agree well within numerical accuracy. As our best estimate of the true threshold, we take the average of the two extrapolated thermodynamic values.

\begin{figure}[t]
    \centering
    \includegraphics[width=.96\linewidth]{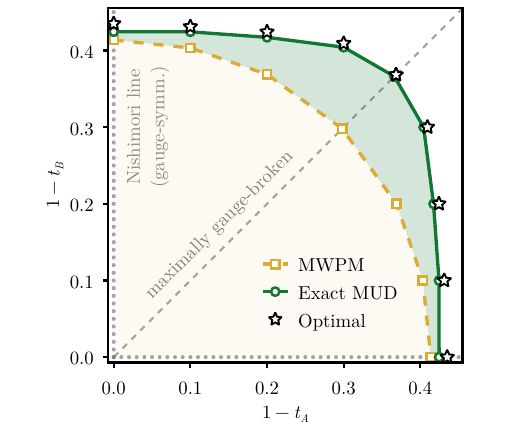}
    \caption{\textbf{Decodability and entanglement threshold phase diagram.} Extrapolated thresholds in the $(t_A,t_B)$ plane for the bare MWPM decoder (yellow squares) and the exact MUD (green circles), together with the decoder-independent entanglement threshold (star markers); see Fig.~\ref{fig:decoding_extrapolation}. The shaded regions mark where the respective decoder recovers long-range order. The green wedge indicates the regime decodable by MUD but not by MWPM. The dotted lines mark the Nishimori line ($t_A=\pi/4$ or $t_B=\pi/4$) and the dashed diagonal ($t_A=t_B$) the case of maximal gauge-symmetry breaking, where the gain of MUD over MWPM is largest.
    }
    \label{fig:mud_phase}
\end{figure}

\section{Finite-size analysis of decoding thresholds}
\label{app.finite_size}
In this appendix, we present the finite-size analysis of the decoding threshold. We observe that the Binder cumulant, even for moderately large systems ($L=20$--$64$), yields threshold values that differ from those reported in the literature for the Nishimori line~\cite{b8y5-k3y6,Mukherjee2026}, which we attribute to finite-size effects and open-boundary contributions. Indeed, when we perform the data collapse for progressively larger systems as in Appendix~\ref{app.sub.finite_size}, we find that the obtained finite-size threshold is roughly a linear function of $1/\d$; see Fig.~\ref{fig:decoding_extrapolation}. Extrapolating this function to $1/\d\to 0$ yields our estimate of the intrinsic decoder's threshold, free of finite-size and boundary corrections, though subject to the uncertainty of the extrapolation.

Surprisingly, we find that the {\sl entanglement threshold} extracted from $I_c$ and  $C^{(2)}_c$ is somewhat smaller than that of the optimal MUD decoder; see Fig.~\ref{fig:mud_phase}. On the $t_A=t_B$ line, the two thresholds agree within the numerical accuracy. It is unclear whether the discrepancy stems from the linear extrapolation (the data for the optimal decoding threshold seem to have a little curvature) or reflects a fundamental separation between the decodability and entanglement thresholds, as in Ref.~\cite{Roser_Decoding_2023}.

\section{Analytic properties of gauge-oblivious decoding}
\label{app.mwpm}

In this Appendix, we show that the leading contribution to the thresholds of the gauge-oblivious MWPM decoder with high accuracy depends only on $A\equiv\sin2t_A\sin 2t_B$. To this end, we compute the decoded expectation value of the two-point correlation function $C_{ij} = \langle\sigma_i\sigma_j\rangle_\text{dec}$ to analyze its parametric dependence on $t_{A/B}$ parameters. Since odd moments $\langle\prod_{i=1}^{2n+1}\sigma_i\rangle$ vanish and the analysis is the same for any even moment $\langle\prod_{i=1}^{2n}\sigma_i\rangle$, the results hold for any observable of $\vect{\sigma}$.

Given the observed syndromes $\vect{s}$, the MWPM decoder outputs the chain $\vect{c}(\vect{f})$, which depends only on the endpoint $\vect{f}=\partial\vect{s}$. Then, the vertex spins inside the closed dual loops $\vect{s}\vect{c}(\vect{f})=\partial\vect{D}(\vect{s})$ are corrected. The decoded correlation function can be computed as 
\begin{align}\label{eq:C1}
    C_{ij} &= 
    \sum_{\vect{\sigma}} \sum_{\vect{s}} \sigma_i\sigma_j D_i(\vect{s})D_j(\vect{s}) P(\vect{\sigma},\vect{s})
    \\&=
    \sum_{\vect{\sigma}} \sum_{\vect{s}} \sigma_i\sigma_j \left(\prod_{l\in {\cal S}_{ij}}s_{l} c_l(\vect{f})\right) P(\vect{\sigma},\vect{s}),
\end{align}
where we expressed the region $\vect{D}(\vect{s})$ inside the loop $\vect{s} \vect{c}$ through $D_i(\vect{s})D_j(\vect{s}) = \prod_{l\in {\cal S}_{ij}}s_{l} c_l(\vect{f})$ the product over an edge chain ${\cal S}_{ij}$ connecting vertices $i$ and $j$. (Recall that $\vect{b} =\partial\vect{D}$ implies that $b_{ij}=D_iD_j$ for any edge $\langle ij\rangle$.) We now exchange the sum over all 1D chains $\vect{s}$ for sums over the endpoints $\vect{f} = \partial\vect{s}$ and all vertex configurations $\vect{r}$. Any chain $\vect{s}$ can be expressed in terms of $\vect{f}$ and $\vect{r}$ as $s_{ij} = [s_f]_{ij}r_i r_j$, where $\vect{s}_f$ is a chain with the same endpoints $\vect{f} = \partial\vect{s}=\partial\vect{s}_f$ \footnote{This parametrization is unique modulo total sign flip $\vect{r}\to-\vect{r}$.}. Without loss of generality, we choose $\vect{s}_f=\vect{c}(\vect{f})$ and obtain
\begin{align}\label{eq:C3}
    C_{ij} &= 
    \frac{1}{2}\sum_{\vect{\sigma}} \sum_{\vect{f},\vect{r}}  \sigma_i\sigma_j  r_i r_j 
    P(\vect{\sigma},\vect{c}(\vect{f})\partial\vect{r})
    \\&= 
    \frac{1}{2}\sum_{\vect{\sigma}'} \sum_{\vect{f}}  \sigma'_i\sigma'_j  
    \sum_{\vect{r}} P(\vect{\sigma}'\vect{r},\vect{c}(\vect{f})\partial\vect{r})\label{eq:C4},
\end{align}
where we defined $\vect{\sigma}'=\vect{\sigma}\vect{r}$. Now, we express the joint probability distribution Eq.~\eqref{eq.P_s_sigma} as
\begin{align}\label{eq:C5}
    P(\vect{\sigma}'\vect{r},\vect{c}\partial\vect{r}) \propto\prod_{\langle ij \rangle} (1+A c_{ij}\sigma'_i \sigma'_j + B c_{ij}r_ir_j),
\end{align}
where $A = \sin 2t_A\sin 2t_B$ and $B = -\cos 2t_A\cos 2t_B$. The $B$ term parametrizes the explicit gauge symmetry breaking and vanishes on the Nishimori line $t_B=\pi/4$. For $B=0$, the gauge symmetry implies $P(\vect{\sigma}'\vect{r},\vect{c}\partial\vect{r})=P(\vect{\sigma}',\vect{c})$.

We then use the high-temperature loop expansion 
\begin{align}\label{eq:loop_expansion}
    \prod_{\langle ij \rangle} (1+K_{ij})
    =
    \sum_{E'\subseteq E}\prod_{\langle ij \rangle\in E'} K_{ij},
\end{align}
where the sum is over all subsets of edges $E'\subseteq E$. To apply the expansion, we factorize the right-hand side of Eq.~\eqref{eq:C5} and  obtain

\begin{align}
    P(\vect{\sigma}'\vect{r},\vect{c}\partial\vect{r})\propto \prod_{\langle ij \rangle} (1+A c_{ij}\sigma'_i \sigma'_j)
    \prod_{\langle ij \rangle} \left(1 + r_ir_j K_{ij}\right)
    \\= 
    \prod_{\langle ij \rangle} (1+A c_{ij}\sigma'_i \sigma'_j)\sum_{E'\subseteq E}\prod_{\langle ij \rangle\in E'} r_ir_j\; K_{ij}, \label{eq:C9}
\end{align}
where $K_{ij} = \frac{B c_{ij}}{1+A c_{ij}\sigma'_i \sigma'_j}$.

Notice that Eq.~\eqref{eq:C4} depends only on the gauge-averaged probability distribution 
\begin{align}\label{eq:C10}
    P_\text{eff}(\vect{\sigma}', \vect{c})= \sum_{\vect{r}} P(\vect{\sigma}'\vect{r},\vect{c}\partial\vect{r}).
\end{align}

We insert Eq.~\eqref{eq:C9} into Eq.~\eqref{eq:C10} and analyze which configuration of $E'$ gives a non-zero contribution to the sum. Any term that has an odd power of $r_i$ vanishes under summation over $r_i=\pm1$. Thus, only if $E'$ is a loop, the product $\prod_{\langle ij \rangle\in E'} r_ir_j = 1$ is trivial and gives a non-zero contribution. Hence, we obtain the averaged probability distribution
\begin{align}
P_\text{eff}(\vect{\sigma}', \vect{c})=
\prod_{\langle ij \rangle} (1+A c_{ij}\sigma'_i \sigma'_j)\sum_{L\subseteq E:\partial L=0} \prod_{\langle ij \rangle\in L} K_{ij}
\\=
\sum_{L\subseteq E:\partial L=0} 
\prod_{\langle ij \rangle\in E \setminus L} (1+A c_{ij}\sigma'_i \sigma'_j)\prod_{\langle ij \rangle\in L} B c_{ij}\,.
\end{align}
We now arrange terms according to their power of $B$, since the parameter $B$ is bounded $|B|\leq1$, and near the strong measurement limit $t_A=t_B=\pi/4$ is small. Therefore, the leading contribution $O(B^0)$ comes from an empty set $L=\emptyset$. The first sub-leading contribution comes from the shortest non-trivial loop. On a square lattice, the shortest possible loops have length $|L|=4$ and thus the corresponding terms have a suppression factor $B^4$.

We now estimate the magnitude of $B=-\cos 2t_A\cos 2t_B$ in the physically relevant regime. Recall that on the Nishimori line $(t_B=\pi/4)$, the threshold is $t^\text{c}_A\approx 0.584 \pi/4$ corresponding to an error rate $p^\text{c}=0.103$. In the region on the $(t_A,t_B)$ plane where MWPM can decode the GHZ state (approximately given by $\sin(2t_A)\sin(2t_B)\geq\sin(2t^\text{c}_A)\approx 0.794$), the parameter is bounded by $B\leq 0.206$, with the bound saturated at $t_A=t_B$, and the first sub-leading term is suppressed by a factor of at most $B^4\leq 1.8\times 10^{-3}$.

If we neglect small corrections due to $B$ in Eq.~\eqref{eq:C5}, we immediately see that all observables depend only on the parameter $A$. In particular, this implies that the threshold depends on $A$ with the first correction of the order $O(B^4)$. Numerically, we find remarkable agreement between the MWPM threshold in the $(t_A,t_B)$ parameter space and the analytic curve $\sin(2t_A)\sin(2t_B)= 0.794=1-2p^\text{c}$.

More generally, for any gauge-oblivious decoder, in the sense that the correction $\vect{c}(\vect{f})$ depends only on the fluxes, the observables can be computed from the distribution averaged over the gauge variables
\begin{align}
P_\text{eff}(\vect{\sigma}',\vect{c})=
\sum_{\vect{r}}P(\vect{\sigma}'\vect{r},\vect{c}\partial\vect{r}),
\end{align}
which corresponds to the Nishimori line ($t_B=\pi/4$) with an effective $t^\text{eff}_A$ satisfying $\sin(2t^\text{eff}_A)=\sin(2t_A)\sin(2t_B)$ (up to the corrections $B^4$).

\section{2D repetition code}\label{app.repetition}

The MUD decoder is designed to maximize the long-range order of the GHZ state parametrized by the magnetization squared. In the context of the repetition code, the GHZ state is the $|+\rangle_\text{L}$ eigenstate of the logical $X_\text{L}$ operator. As in measurement-based state preparation, bit-flip noise introduces domain walls that are identified by measuring syndromes---parity checks on the edges between two data qubits. 

When the syndrome measurements are perfect $t_A=t_B=\pi/4$, the domain walls can always be eliminated, and the logical information is preserved if fewer than half of the data qubits are flipped by the noise, i.e., the bit-flip error rate is $p_\text{x}<0.5$. However, away from the fine-tuned strong-measurement point $t_A,t_B<\pi/4$, two distinct situations arise: the 1D repetition chain and the 2D repetition lattice. In the former case, the syndromes are not redundant. Any syndrome error propagates uncontrollably, so there is no finite threshold, $t^\text{c}_A=t^\text{c}_B=\pi/4$. In the 2D case, any parity check $s_{ij}$ between qubits $i$ and $j$ can be independently verified by any chain of syndromes ${\cal S}_{ij}$ with endpoints $i$ and $j$. This redundancy results in a finite threshold $t^\text{c}_A,t^\text{c}_B<\pi/4$.

As in the GHZ-state preparation, the data and syndrome qubits in the weak-measurement 2D repetition code are governed by the RBIM-like distribution 
\begin{align}
    P(\vect{\sigma},\vect{s})\propto e^{\beta \left[\sum_{\langle ij\rangle}\left(J_{s_{ij}}\sigma_i \sigma_j +hs_{ij}\right)+ \sum_i h_x  \sigma_i\right]}  ,
\end{align}
where $\beta h_x$ controls the strength of the bit-flip noise.  The statistics of the measurement-based protocol are recovered when $p_x=0.5$ and $h_x=0$.

\subsection{Maximum likelihood decoder}

In the repetition code, the logical information is preserved if, after the correction, the majority of the data qubits point in the same direction as the original logical state. For concreteness, we assume that the logical state is $|1\rangle_L$, and the logical information is restored by the decoder if $\langle \hat{M}_z \rangle_\text{dec}>0$. Given the observed syndromes $\vect{s}$, the MLD decoder chooses the correction $\vect{c}$ that is most likely (with respect to $P(\vect{\sigma}|\vect{s})$) to result in $\langle \hat{M}_z \rangle_\text{dec}>0$. The probability that the correction $\vect{c}$ recovers the logical information given $\vect{s}$ is 
\begin{align}
    P_\text{L}(\vect{c}|\vect{s}) = \sum_{\vect{\sigma}: \left(\sum_{i}\sigma_i \tau_i\right)>0}P(\vect{\sigma}|\vect{s}),
\end{align}
where $\tau_i=\tau_i(\vect{c}\vect{s})$ is the {\sl smallest} region whose boundary is a loop $\vect{c}\vect{s}$, i.e., $\tau_i\tau_j=c_{ij}s_{ij}$. The maximum-likelihood decoder then chooses the most probable correction 
\begin{align}
    \text{MLD: }\vect{s} \to \vect{c} = \mathop{\operatorname{argmax}}\limits_{\vect{c}} P_\text{L}(\vect{c}|\vect{s}).
\end{align}

\subsection{Maximum-utility decoder}

The MLD for the 2D repetition code can be formulated as MUD with expected utility given by the logical recovery probability
\begin{align}
    P_\text{L}(\vect{\tau}|\vect{s}) = \sum_{\vect{\sigma}} \mathcal{U}(\vect{\tau},\vect{\sigma}) P(\vect{\sigma}|\vect{s}),
\end{align}
and the utility function 
\begin{align}
    {\cal U}_\text{MLD}(\vect{\tau},\vect{\sigma}) = \theta\left(\sum_{i}\sigma_i \tau_i\right),
\end{align}
where $\theta(x)$ is a Heaviside step function. This choice of utility function is analytically intractable due to the nonlinearity of the $\theta$-function. 

Another natural choice of the utility function corresponds to magnetization 
\begin{align}
    {\cal U}_M(\vect{\tau},\vect{\sigma}) = \frac{1}{N}\sum_{i}\sigma_i \tau_i. 
\end{align} 
In this case, the expected utility is 
\begin{align}
    \mathbb{E}[{\cal U}_M] =  \frac{1}{N}\sum_{i} \tau_i \langle \sigma_i \rangle_{\vect{s}}.
\end{align}
The solution for $\vect{\tau}$ that maximizes the expected utility is then $\tau_i = \operatorname{sign}(\langle \sigma_i \rangle_{\vect{s}})$. This choice of utility function corresponds to correcting the most-likely error on each qubit individually. In Bayesian statistics, it is known as a (qu-)bitwise maximum a posteriori (MAP) decoder. By contrast, MLD is a blockwise MAP decoder that maximizes the probability of correcting all qubits simultaneously. 

\bibliography{ref}

\end{document}